\documentclass[useAMS,usenatbib,twocolumn,preprintnumbers,nofootinbib]{revtex4}
\usepackage{graphicx}% Include figure files 
\usepackage{dcolumn}% Align table columns on decimal point 
\usepackage{bm}% bold math 
\usepackage{epsfig} 
\usepackage{amsfonts}
\usepackage{amsmath}
\usepackage{amssymb}
\usepackage[usenames]{color}
\usepackage[colorlinks=true]{hyperref}

\hypersetup{colorlinks,citecolor=blue,linkcolor=blue,urlcolor=blue}
\hypersetup{final=true}

\begin{document}

\title{Is there evidence for a hotter Universe?}

\author{Carlos A. P. Bengaly$^{1}$\email{carlosbengaly@on.br}, Javier E. Gonzalez$^{2}$\email{gonzalezjavier@gmail.com}, Jailson S. Alcaniz$^{1,2,3}$\email{alcaniz@on.br}}
\affiliation{$^1$Observat\'orio Nacional, 20921-400, Rio de Janeiro - RJ, Brazil}
\affiliation{$^2$Departamento de F\'{\i}sica, Universidade Federal do Rio Grande do Norte, 59072-970, Natal, RN, Brazil}
\affiliation{$^3$Instituto Nacional de Pesquisas Espaciais/CRN, 59076-740, Natal, RN, Brazil}

\date{\today}

\begin{abstract}

The measurement of present-day temperature of the Cosmic Microwave Background (CMB), $T_0 = 2.72548 \pm 0.00057$ K (1$\sigma$), made by the Far-InfraRed Absolute Spectrophotometer (FIRAS) as recalibrated by the Wilkinson Microwave Anisotropy Probe (WMAP), is one of the most precise measurements ever made in Cosmology. On the other hand, estimates of the Hubble Constant, $H_0$, obtained from measurements of the CMB temperature fluctuations  assuming the standard $\Lambda$CDM model exhibit a large ($4.1\sigma$) tension when compared with low-redshift, model-independent observations. Recently, some authors argued that a slightly change in $T_0$ could alleviate or solve the $H_0$-tension problem. Here, we investigate evidence for a hotter or colder universe by performing an independent analysis from currently available temperature-redshift $T(z)$ measurements. Our analysis (parametric and  non-parametric) shows a good agreement with the FIRAS measurement and a discrepancy of $\geq 1.9\sigma$ from the $T_0$ values required to solve the $H_0$ tension. This result reinforces the idea that a  solution of the $H_0$-tension problem in fact requires either a better understanding of the systematic errors on the $H_0$ measurements or new physics. 

\end{abstract}

\maketitle

%--------------------------------------------------------------------------------------------------------------------------
\section{Introduction}\label{sec:intro}

About three decades ago, the frequency spectrum of the Cosmic Microwave Background (CMB) radiation was measured by the Far-InfraRed Absolute Spectrophotometer (FIRAS)~\citep{{mather}}. Over a large range of frequencies, the spectrum obtained was an almost perfect blackbody at a temperature $T_0 \simeq 2.73$ K, which is the best blackbody spectrum ever measured. Later on, the FIRAS data were recalibrated using the Wilkinson Microwave Anisotropy Probe observations, resulting in one of the most precise measurements in Cosmology~\citep{fixsen09} (henceforth F09)
\begin{equation}\label{eq:T0_firas}
T_0 = (2.72548 \pm 0.00057)\, \mathrm{K} \;\; \mbox{($1\sigma$)} \,. 
\end{equation}
More recently, measurements of the temperature fluctuations of the CMB across the sky have been used to provide stringent constraints on the other cosmological parameters, such as the Hubble constant~\citep{planck18}
\begin{eqnarray}\label{eq:P18_bestfit}
H_0 &=& (67.36 \pm 0.54) \, \mathrm{km \, s}^{-1} \, \mathrm{Mpc}^{-1}\, \;\; \mbox{($1\sigma$)}  \,, 
\end{eqnarray}
a value that {was} obtained assuming a flat $\Lambda$ - Cold Dark Matter ($\Lambda$CDM) model from the 2018 dat release of the Planck Collaboration (hereafter P18). Other cosmological probes, such as distance measurements from Type Ia Supernovae~\citep{scolnic18} and the baryonic acoustic oscillation (BAO) signal from galaxy clustering observations~\citep{icaza-lizaola19} have also confirmed the description of the universe provided by the $\Lambda$CDM model.

In spite of the remarkable concordance among the estimates and measurements of the standard model parameters from different probes, the Planck estimate of the Hubble Constant exhibits a $4.1\sigma$ tension with measurements of the current expansion rate from low-redshift standard candles, which were obtained in a model independent-way by the SH0ES experiment~\citep{Ried19,riess18}, 
\begin{equation}
H_0 = (73.5 \pm 1.4) \, \mathrm{km \, s}^{-1} \, \mathrm{Mpc}^{-1} \;\; \mbox{($1\sigma$)}  \,. 
\end{equation}
Such a large tension is not easily reconciled with extensions of the standard cosmology, even though several theoretical attempts have been proposed. In some case, they are not able to satisfactorily explain the $H_0$ tension without creating additional discrepancies with the measurements of other parameters~(see e.g. \cite{riess19} and references therein).

Another possible route to explain this tension consists in revising the fundamental prior assumptions in cosmological measurements, as recently done by~\cite{ivanov20} (henceforth IAL20). In their analysis, $T_0$ was indirectly estimated from Planck observations using gravitational lensing and the integrated Sachs-Wolfe effect. However, in order to solve the $H_0$-tension problem, the value of $T_0$ obtained was much smaller than the FIRAS measurement, which amounts to saying that the $H_0$ tension was replaced by a $T_0$-tension problem between the Planck and FIRAS measurements of the present-day CMB temperature. 

A step further was given by~\cite{bose20} (hereafter BL20) who not only relaxed the $T_0$ constraint, but also the assumption of a flat Universe. They found that a hotter and open Universe could indeed alleviate the $H_0$ tension in a significant way, in addition to milder tensions currently present in the $\Lambda$CDM model, like the low CMB quadrupole value, the  higher CMB lensing amplitude, and the conflicting value of the normalization of the matter power spectrum on scales of 8$h^{-1}$ Mpc ($\sigma_8$) between measurements from CMB and galaxy surveys. 

Motivated by the IAL20 and BL20 analyses, in what follows we perform an independent estimate of $T_0$ directly from the current temperature-redshift data, $T(z)$. Our goal is to verify whether independent temperature measurements support a slightly colder or hotter universe today, in agreement with the IAL20 and BL20 results, respectively. The $T(z)$ data used in our analysis were obtained from observations of the Sunyaev-Zeldovich (SZ) effect towards clusters~\citep{luzzi09, saro14, hurier14, luzzi15, demartino15}. We perform a parametric fit and a non-parametric regression analysis of the $T(z)$ data and obtain an interval of values of $T_0$ consistent with the FIRAS measurement and at least $2.8\sigma$ ($1.9\sigma$) off from the value required by the IAL20 (BL20) estimate to solve the $H_0$ and other cosmological tensions.

%%%%%%%%%%%%%%%%%%%%%%%%%%%%%%%%%%%%%%%%%%%%%%%%%%%%%%

\section{Analysis}

The  primary data set used in this analysis consists of 103 SZ measurements  within the redshift interval $0.01<z<0.97$~\citep{luzzi15}. This sample was obtained from the Planck catalog that comprises 861 confirmed galaxy clusters, 816 of them having known redshifts. This data set is a selection of galaxy clusters with X-ray and optical information with high quality (S/N$\geq$ 6), which allows to determine  $T_{\rm CMB}(z)$ estimates from individual clusters with a precision of up to 3$\%$. 

Additionally, we also consider two other combinations of $T(z)$ measurements compiled by~\cite{avgoustidis16} and \cite{hurier14}, respectively. Combination 1 (henceforth  comb1) corresponds to 12 $T(z)$ measurements within the range $0.13<z<1.02$~\citep{saro14}, along with 18  $T(z)$ measurements in the interval $0.03<z<0.97$~\citep{hurier14}. Combination 2 (henceforth comb2) consists on 13 $T(z)$ measurements in the range $0.02<z<0.55$~\citep{luzzi09} combined with the $T(z)$ measurements by~\cite{hurier14}. 

As is well known, in the absence of cosmic opacity or photon non-conservation, the temperature evolution law of the CMB is given by
\begin{equation}\label{eq:tz}
T(z) = T_0(1+z) \,.
\end{equation}
Such result does not depend on cosmology,
and no significant deviation from this law has been found using different compilations of $T(z)$ measurements (see. e.g.~\cite{Lima00, demartino12, luzzi15, demartino15, avgoustidis16, arjona20}). 

Initially, we obtain our estimate of the present CMB temperature using a parametric fitting procedure, where $T_0$ is estimated from $T(z)$ observations through a usual $\chi^2$ minimization. 
Our prior choice  is $T_0 [\mathrm{K}]: \mathcal{U}(2.4,3.1)$, where $\mathcal{U}(a,b)$ represents an uniform (flat) prior ranging from $a$ to $b$. We compare our $T_0$ estimates with the following results:
\begin{eqnarray}\label{eq:T0} \nonumber
T_0 & = & (2.72548 \pm 0.00057)\, \mathrm{K} \; \mbox{(F09)}\\ \nonumber
T_0 & = & (2.564^{+0.049}_{-0.051})\, \mathrm{K} \; \mbox{(IAL20, P18+SH0ES)}\\ \nonumber
T_0 & = & (2.839 \pm 0.046)\, \mathrm{K} \; \mbox{(BL20, P18+SH0ES+BAO)}
\end{eqnarray}
The latter two estimates correspond to the final values of $T_0$ obtained by IAL20 and BL20, respectively.

%%%%%%%%%%%%%%%%%%%%% Fig 1
\begin{figure*}[!ht]
\centering
\includegraphics[width=0.49\linewidth, height=7.3cm]{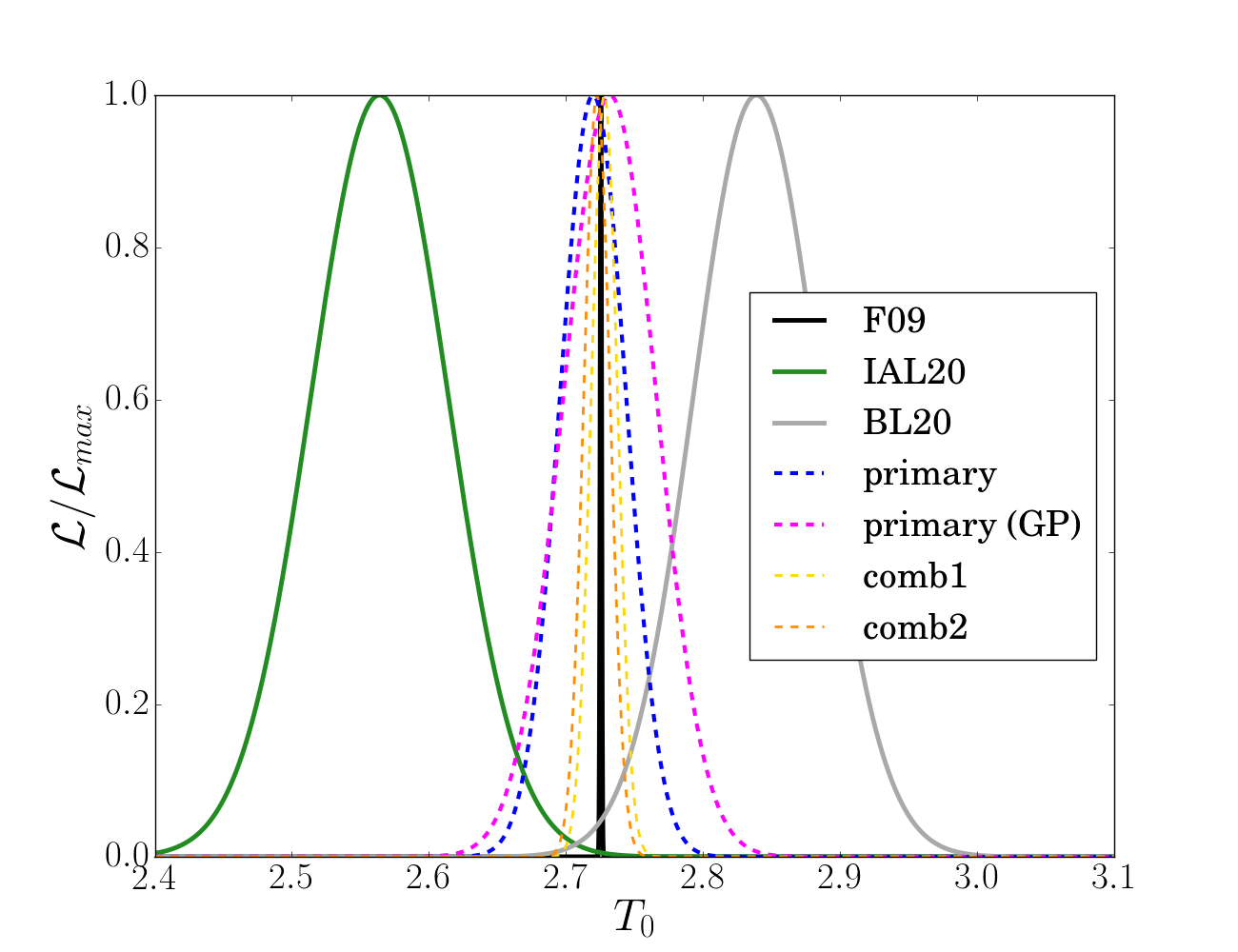}
\includegraphics[width=0.49\linewidth, height=7.3cm]{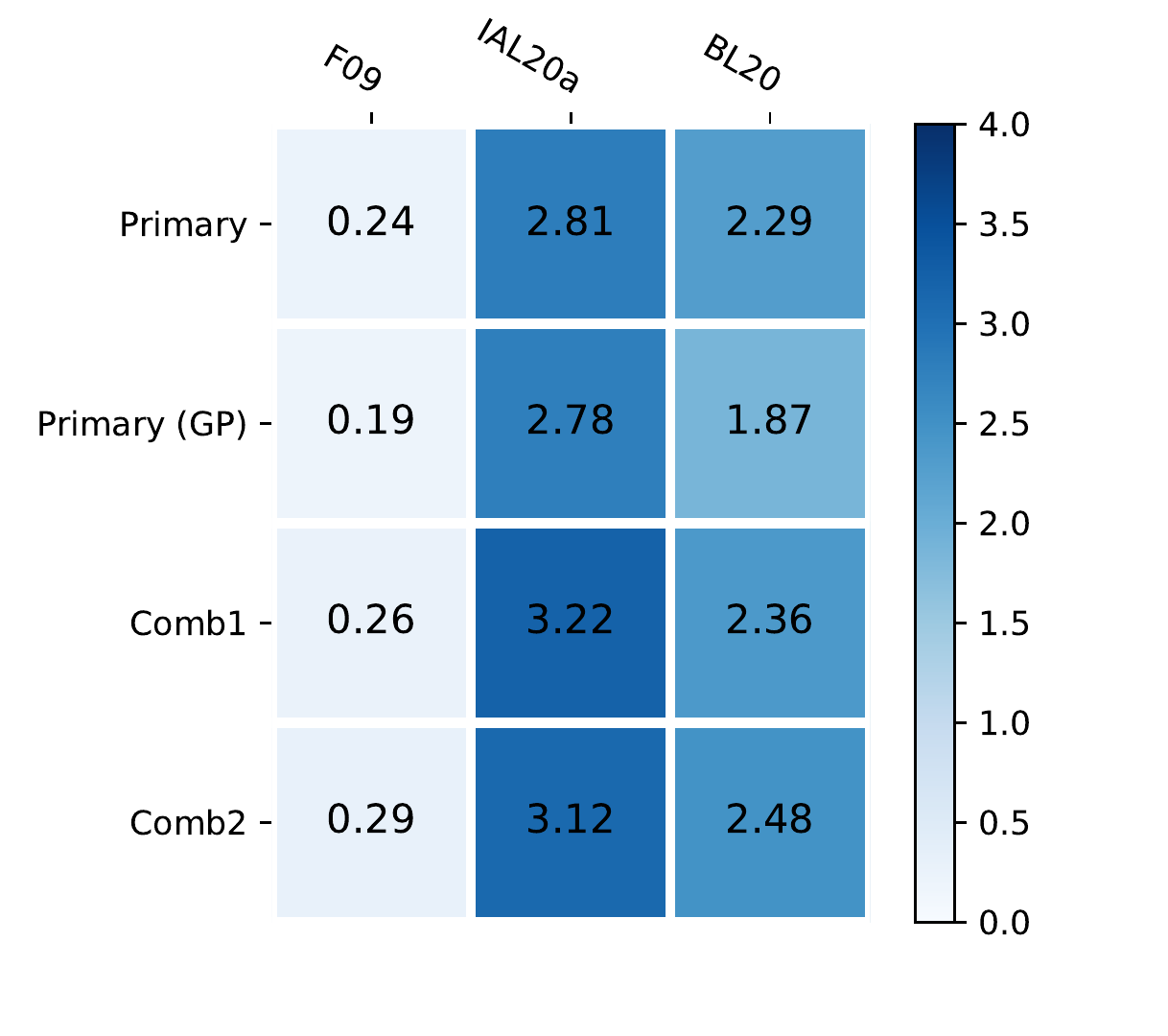}
\includegraphics[width=0.49\textwidth, height=6.4cm]{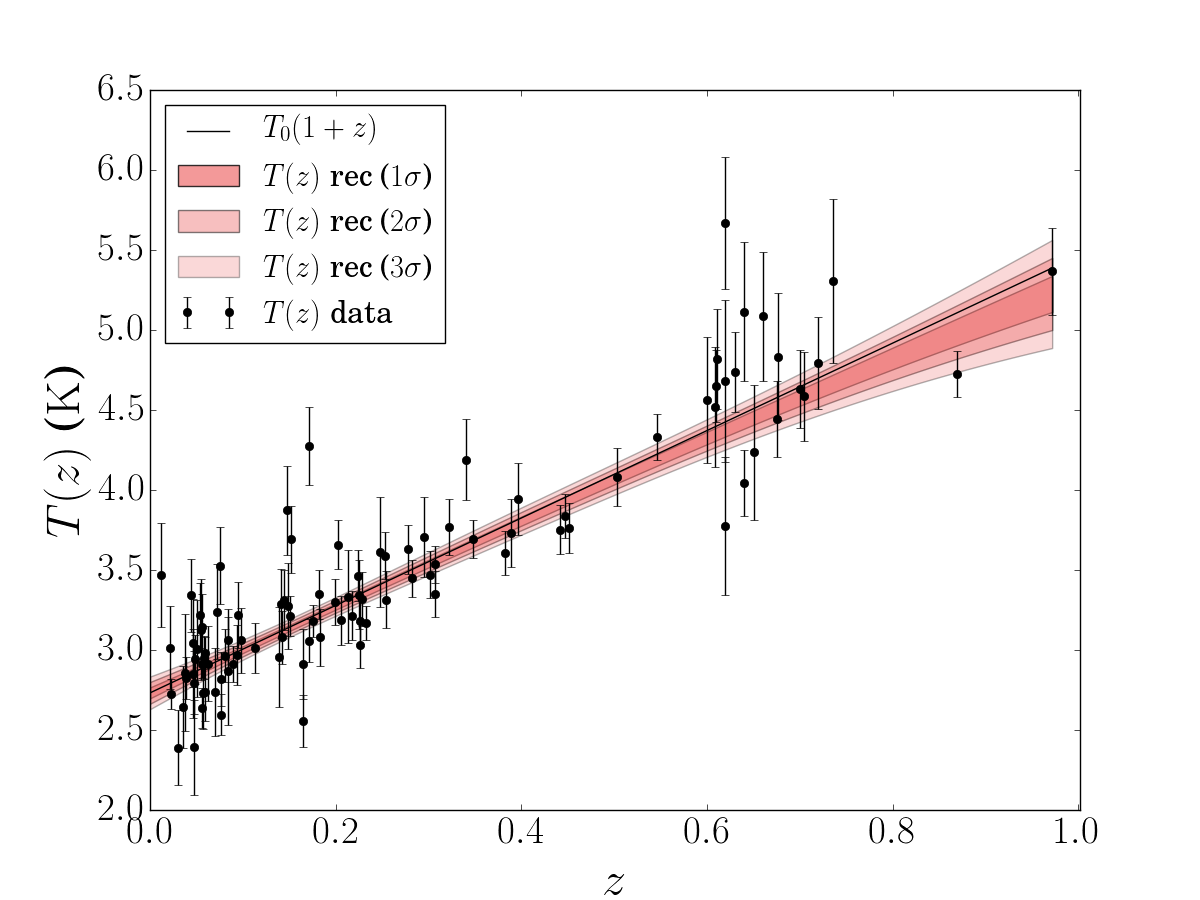}
\label{fig:likeh_heatmap}
\caption{{\it Upper left panel}: The normalized $T_0$ likelihoods of F09 (black), IAL20 (dark green), BL20 (dark gray), plotted against the $T_0$ measurements obtained for each data-set (primary in blue, comb1 in golden, comb2 in dark orange). We also plot the result obtained for the primary data-set using GP, as displayed in magenta {\it Upper right panel}: A heatmap of the tension between F09, IAL20 and BL20 measurements with our estimates, as calculated using Eq.~(\ref{eq:tension}). {\it Bottom panel}: The GP reconstructed curves at $1$, $2$, and $3\sigma$ from the $T(z)$ measurements of our primary data set. We also plot the standard law $T(z)=T_0(1+z)$ for the sake of comparison.}
\end{figure*}
%%%%%%%%%%%%%%%%%%%%% Fig 1

% %%%%%%%%%%%%%%%%%%%%% Fig 2
% \begin{figure*}[!ht]
% \centering
% \includegraphics[width=0.49\textwidth, height=6.4cm]{Tz_primary_reconst_SqExp.png}
% % \includegraphics[width=0.49\textwidth, height=6.4cm]{Tz_primary_reconst_SqExp_zoom.png}
% \caption{The GP reconstructed curves at $1$, $2$, and $3\sigma$ from the $T(z)$ measurements of our primary data set. We also plot the standard law $T(z)=T_0(1+z)$ for the sake of comparison.}
% \label{fig:Tz}
% \end{figure*}
% %%%%%%%%%%%%%%%%%%%%% Fig 2

The discrepancy between these measurements and our estimate is quantified by the following estimator
\begin{equation}\label{eq:tension}
\mathcal{T} = \frac{|T_{0,{\rm{exp1}}}-T_{0,{\rm exp2}}|}{\sqrt{\sigma_{T_{0,{\rm{exp1}}}}^2 + \sigma_{T_{0,{\rm{exp2}}}}^2}} \;,
\end{equation}
where $T_{0,{\rm{exp1}}}$ and $T_{0,{\rm{exp2}}}$ represent $T_0$ measurements obtained by two different experiments or data-sets. 

%%%%%%%%%%%%%%%%%%%%%%%%%%%%%%%%%%%%%%%%%%%%%%%%%%%%%%

\section{Results}

Table~\ref{tab:T0_measurements} shows our $T_0$ estimates for our primary data set, comb1 and comb2. We find that they are consistent with each other at a $2\sigma$ confidence level (see also the {upper left panel} of Fig.~(1)). %\ref{fig:likeh_heatmap}. 
In order to quantify these results, we also present the tension $\mathcal{T}$ between the different $T_0$ estimates in a heatmap displayed in the {upper right plot} of Fig.~(1). %\ref{fig:likeh_heatmap}. 
All our estimates are in very good agreement with F09, i.e., $\mathcal{T}=(0.2-0.3)\sigma$, but they show discrepancy with BL20, such as $\mathcal{T}=(1.9-2.4)\sigma$, and some tension with IAL20, which ranges between $\mathcal{T}=(2.8-3.2)\sigma$.

\begin{table}[!ht]
    % \centering
    \begin{tabular}{cccccc}
    \hline\hline
      Data-set & $T_0$(K) & {$\chi_{\rm red}^2$} \\ \hline\hline
      primary & $2.7198 \pm 0.0241$ & {$1.67$} \\
      primary (GP) & $2.7320 \pm 0.0339$ & {-}\\
      comb1 & $2.7280 \pm 0.0096$ & {$0.87$} \\
      comb2 & $2.7226 \pm 0.0096$ & {$0.92$} \\
      \hline\hline
    \end{tabular}
    \caption{Estimates of $T_0$ obtained for each data-set at {$1\sigma$ confidence level}, {along with the reduced $\chi_{\rm red}^2$ value, i.e., $\chi^2$ divided by the number of degrees of freedom}.}
    \label{tab:T0_measurements}
\end{table}

For completeness, we also perform a non-parametric reconstruction analysis of our primary data set.  We apply Gaussian Processes and use the {\sc GaPP} code~(see \cite{seikel12} and references therein)\footnote{GaPP is available at \url{https://github.com/carlosandrepaes/GaPP}.}. Since this is a model-independent analysis, we can obtain $T_0$ regardless of the standard model assumption of adiabaticity, by the same token of measuring $H_0$ with $H(z)$ measurements~\citep{gomezvalent18, bengaly20}. We find a consistent result with the value of $T_0$ derived from our parametric fit - albeit with larger uncertainty. The reconstructed curves of $T(z)$ ($1\sigma$ to $3\sigma$) obtained with the primary data set are shown in the {lower plot of Fig.~(1)}. The GP results are in full agreement with the F09 measurement with $\mathcal{T}=0.19\sigma$, and show a discrepancy of $\mathcal{T}=2.78\sigma$ and $\mathcal{T}=1.87\sigma$ with the IAL20 and BL20 estimates, respectively. It is worth mentioning that these result were obtained assuming the Squared Exponential GP kernel, but we also verified that changing the kernel does not yield appreciably different results.

\section{Conclusions}

The tension between $H_0$ measurements from early and late Universe probes is currently a matter of great debate and controversy. The existence of milder tensions, such as the value of $\sigma_8$, the CMB lensing amplitude, and the low CMB quadrupole, hint at a possible departure from the standard cosmological model. Several approaches have been proposed to solve it, but none of them could successfully address all these issues. The possibility of a slightly colder and flat or hotter and open universe, however, can alleviate most of these tensions, as shown by IAL20 and BL20, respectively. This is particularly interesting because it does not require a profound reformulation of the standard cosmology.

In this paper, we performed an independent analysis and  estimated $T_0$ using different combinations of $T(z)$ measurements obtained from the SZ effect~\citep{luzzi15}. The analysis performed is model-independent and furnishes $T_0$ estimates  in very good  agreement with the FIRAS measurement. On the other hand, our results from both a parametric fit and non-parametric reconstruction show only marginal evidence for a hotter or colder universe that could reconcile the current cosmological discrepancies and tensions, as discussed by IAL20 and BL20. 

This may be understood as an evidence that a possible solution to the $H_0$ tension in fact requires  a better understanding of the systematic errors on $H_0$ measurements or a further exploration of physics beyond the standard cosmological model. 

\vspace{0.2cm}

{\it Acknowledgments:} We thank Gemma Luzzi and Alex Saro for sending the $T(z)$ data compilation reported in \cite{luzzi15} and \cite{saro14}, respectively. CB acknowledges financial support from the Programa de Capacita\c{c}\~ao Institucional PCI/ON/MCTI. JEG acknowledges financial support from the Coordenação de Aperfeiçoamento de Pessoal de Nível Superior (CAPES). JSA is supported by CNPq (Grant no. 310790/2014-0) and Funda\c{c}\~ao de Amparo \`a Pesquisa do Estado do Rio de Janeiro, FAPERJ (grant no. 233906).

\end{document}